# How Large Should the QM Region Be in QM/MM Calculations? The Case of Catechol O-methyltransferase

*Heather J. Kulik[†,§,%], Jianyu Zhang[¶], Judith P. Klinman[¶], and Todd J. Martinez[†,§,*]*

[†]Department of Chemistry and PULSE Institute, Stanford University, Stanford, CA 94305
[§]SLAC National Accelerator Laboratory, Menlo Park, CA 94025
[¶]Departments of Chemistry and of Molecular and Cell Biology and California Institute for Quantitative Biosciences, University of California, Berkeley, CA 94720

*Email: todd.martinez@stanford.edu

**Abstract:** Hybrid quantum mechanical-molecular mechanical (QM/MM) simulations are widely used in studies of enzymatic catalysis. Until recently, it has been cost prohibitive to determine the asymptotic limit of key energetic and structural properties with respect to increasingly large QM regions. Leveraging recent advances in electronic structure efficiency and accuracy, we investigate catalytic properties in catechol O-methyltransferase, a representative example of a methyltransferase critical to human health. Using QM regions ranging in size from reactants-only (64 atoms) to nearly one-third of the entire protein (940 atoms), we show that properties such as the activation energy approach within chemical accuracy of the large-QM asymptotic limits rather slowly, requiring approximately 500-600 atoms if the QM residues are chosen simply by distance from the substrate. This slow approach to asymptotic limit is due to charge transfer from protein residues to the reacting substrates. Our large QM/MM calculations enable identification of charge separation for fragments in the transition state as a key component of enzymatic methyl transfer rate enhancement. We introduce charge shift analysis that reveals the minimum number of protein residues (ca. 11-16 residues or 200-300 atoms for COMT) needed for quantitative agreement with large-QM simulations. The identified residues are not those that would be typically selected using criteria such as chemical intuition or proximity. These results provide a recipe for a more careful determination of QM region sizes in future QM/MM studies of enzymes.

[%] Present Address: Department of Chemical Engineering, Massachusetts Institute of Technology, Cambridge, MA 02139

## 1. Introduction

A firm understanding of how enzymes facilitate chemical reactions is key for designing molecular catalysts[1] and novel enzymes.[2] Atomistic simulations of enzymes[3] can provide valuable insight distinguishing rate enhancements due to static, local transition-state stabilization[4] from more non-local effects.[5] However, there remains considerable uncertainty regarding the role of the greater protein.[6-10] Enzyme simulation requires a balance of sufficient accuracy to describe chemical rearrangements and catalytic enhancement with low computational cost to enable extensive sampling. Typically, this balance is achieved through a multilevel approach,[11-17] wherein the region of primary interest is treated quantum mechanically (QM), while the surrounding portion of the enzyme is described with an empirical molecular mechanics (MM) model. Largely because of computational limitations, typical QM region sizes (i.e. ligands and a few direct residues) are on the order of tens of atoms.[18-20] There has been much work[17,21-30] to minimize QM/MM boundary effects that might be of concern with small QM regions and to evaluate[31] how advanced, i.e., polarizable,[32-33] force field treatments may improve QM/MM descriptions. However, the requirement to treat crucial[34-35] charge transfer across the QM/MM boundary indicates that boundary-effect minimization and force field adjustment may be insufficient to address the shortcomings of small QM/MM calculations.

Recent advances[34,36-43] in computational efficiency enable fully *ab initio*, quantum chemical simulation of polypeptides[36] as well as QM/MM treatments of enzymes using *ab initio* QM methods and large (more than 100 atoms) QM regions. At the same time, advances in the accurate treatment of exchange within range-separated hybrids in density functional theory (DFT) have led to first-principles methods that can reach quantitative agreement with experiment even for nonreactive problems where carefully fit force fields were once thought to be superior (e.g., in properties of water[44]). Despite these advances, studies of the extent to which quantum effects (e.g. due to polarization or charge transfer) are relevant in enzyme catalysis beyond a small active site region have largely been restricted to semi-empirical QM/MM methods due to computational cost.[45-46] Thus far, *ab initio* QM/MM convergence studies have reported disappointingly slow approach to asymptotic limits for NMR shieldings,[47-48] solvation effects,[49] barrier heights,[50-51] excitation energies,[52] partial charges,[53] and redox potentials.[54] Computational considerations have restricted these studies to i) focus on convergence properties in the context of single point energies of one or few structures[41,47-48,55] and ii) employ local,[53] semi-local,[50,56-59] or global hybrid[48,54] exchange correlation (xc) functionals in DFT. Although it has been possible to carry out one-shot, single point energies of very large systems for some time,[60-66] systematic transition state determination and intermediate geometry optimization, which can require thousands of such single point energies, has been addressed in very few[50,56] QM/MM convergence studies. Additionally, all but one[52] of these studies has been carried out with DFT xc functionals that lack asymptotically correct exchange and produce well-known errors[67-71] in energetics[68,72-76] that likely increase with system size. Thus the extent to which slow QM/MM region convergence is a consequence of errors in the xc approximation versus the result of an increasingly complete treatment of the chemical environment is still unknown.

In this work, we harness recent advances in computational efficiency with asymptotically-correct DFT xc functionals to investigate the convergence of key catalytic properties with increasing QM region size using the model system, catechol *O*-methyltransferase (COMT).[77] COMT regulates neurotransmitters in the human body by transferring a methyl group of *S*-adenosyl-L-methionine (SAM) to deprotonated catecholamines[78] such as dopamine. Methyl transfer is a critical reaction for human health both in neurotransmitter regulation by COMT and in the related DNA methyltransferases[79-80] that regulate gene expression. Unlike in the metalloenzymes that have been the near-exclusive focus of previous QM/MM convergence studies,[50,54,56-59] the reacting substrates in COMT are not covalently bound to the protein. Thus COMT represents a unique opportunity to decouple boundary effects on QM/MM convergence from effects due to increasingly complete treatments of charge transfer and polarization.

All available COMT crystal structures indicate an unusually short SAM C to catecholate O non-bonded distance of 2.45-2.8 Å, in disagreement with classical molecular dynamics simulation[81-82] and most previous QM/MM simulations[7,83-84] which were limited to small QM regions. The crystal structure also indicates strong bidentate coordination of catecholate to an active site $Mg^{2+}$ that is known from biochemical observations to be essential[77,85] for COMT function. However, previous QM/MM simulations have predominantly indicated a preference for weaker monodentate[86-88] or no coordination[7,84] between $Mg^{2+}$ and catecholate. One possible cause for these discrepancies is the frequent exclusion of $Mg^{2+}$ from the QM region.

Thus, COMT provides i) a valuable test case for enlarging present understanding of region-size sensitivity in QM/MM simulations and ii) a representative model enzyme for which electrostatics and charge transfer are expected to be mechanistically critical but are still not well understood. We now systematically determine how properties of reacting substrates at the active site of COMT such as energetics, partial charges, and structural properties approach asymptotic limits with increasingly expansive quantum-mechanical descriptions in QM/MM simulations. This allows us to address both methodological questions about QM region sizes in QM/MM and mechanistic questions about COMT reactivity. The outline of the rest of this work is as follows.

In Sec. 2, we summarize the computational details, and, in Sec. 3, we outline our QM/MM convergence approach. In Sec. 4, we present Results and Discussion on i) the convergence of key properties with increasingly large QM/MM models, ii) mechanistic insight afforded by large QM/MM models, and iii) an approach for systematic determination of atom-economical QM regions in QM/MM calculations. Finally, in Sec. 5, we provide our conclusions.

**2. Computational Details**

Simulations start from the crystal structure of the soluble, human form of COMT[89] (PDB ID: 3BWM), which has been solved in the presence of a dinitrocatecholate (DNC) inhibitor. Six residues at the C terminus and one residue at the N terminus of the protein are unresolved in the crystal structure, producing a 214-residue, 3419-atom model, where the first and last resolved residues were treated as the N and C terminus respectively during preparation by the tleap utility. As a starting point for simulations, we converted DNC to a catecholate anion substrate in the COMT structure by removing the nitro groups, and we preserved three resolved buried water molecules (of 110 total crystal waters resolved in 3BWM, the remainder of which were adjacent to the external surface of the protein) near the catalytically relevant $Mg^{2+}$ ion (HOH411, HOH402, and HOH403 in 3BWM). All other external water molecules were later replaced during solvation of the complete protein. The protein was protonated using the H++ webserver[90-92] assuming a pH of 7.0, which yielded a holoenzyme net charge of -6. Neutralizing $Na^+$ charges were added using the AMBER tleap program.[93] Counterions introduced to produce a neutral simulation cell were always treated with the force field. Generalized Amber Force Field (GAFF) parameters were determined for both SAM and catecholate using the Antechamber code in AMBER for use alongside the ff12SB force field for the rest of the protein in MM simulations.[93]

For MM simulations, a truncated octahedron with 15Å buffer of water from the edge of the protein before NPT equilibration was employed with periodic boundary conditions.

Prior to QM/MM simulations, a well-equilibrated MM structure was obtained as follows: i) 1000 constrained-protein (i.e. only solvent and ions are minimized while the protein is held fixed) and 1000 free-protein minimization steps (i.e. everything in the system is minimized), ii) 20 ps quick NVT heating to 300 K, iii) 5 ns NPT equilibration (p = 1 bar, T = 300 K), and iv) 100 ns of NVE production runs. A representative snapshot was selected from the MM production run by choosing a random structure with C-O distance equal to the mode of the C-O SAM-catechol distance distribution (~3.11 Å)[82] for subsequent AMBER-driven QM/MM geometry optimizations and nudged elastic band[94] calculations.

For all QM/MM simulations, we carried out combined quantum mechanical (QM) and molecular mechanics (MM) calculations using our TERACHEM package[95] for the QM portion and AMBER 12[93] for the MM component. The QM region is modeled with DFT using the range-separated exchange-correlation functional ωPBEh (ω=0.5 bohr$^{-1}$) with the 6-31g[96] basis set, a combination we have previously benchmarked for protein structure.[36] In the QM/MM calculations, an aperiodic spherical droplet was extracted from the production MM results by selecting the largest radius (at least 10 Å of solvent) that could be inscribed in the truncated octahedron using the center of mass utility in PyMOL.[97] Comparisons to results obtained by directly starting QM/MM calculations from the crystal structure geometry are provided in the Supporting Information. Voronoi deformation density (VDD) charges[98] were chosen to assess intersubstrate and substrate-protein charge transfer due to their relatively low basis set sensitivity.[98]

## 3. Approach

QM regions were obtained by starting from a model that consisted of only SAM and catecholate substrates and identifying residues that were within increasing cutoff distances from these reactants. We chose a total of 10 QM region sizes for QM/MM calculations ranging from the reactants-only (including the $Mg^{2+}$ ion) model **1** (64 atoms and 0 protein residues in the QM region) to a largest model **10** consisting of 940 atoms (reactants and 56 protein residues in the QM region) (Table 1). Regions were chosen by sequentially increasing the cutoff distance at values of 0.00, 1.75, 2.00, 2.25, 2.50, 2.75, 3.00, 5.00, 6.00, and 7.00 Å from any atom in a residue to any atom in either SAM or catecholate, as determined using determined using distance functions in PyMOL[97] on the crystal structure (Figure 2). These distance cutoffs were chosen to obtain region sizes that differed by around 5 residues in size for the small- to mid-sized QM regions (the list of residues in each QM region is provided in Supporting Information Table S1). In the two largest regions, two charged residues (D150 and D205) were excluded despite satisfying the distance cutoffs in order to obtain a QM region with a net charge of -1 rather than -3, avoiding challenges for DFT with highly charged anionic systems.[99-101] Notably, D150 and D205 were not covalently bound to any other residues in the QM region, and their exclusion thus reduced the number of covalent bonds spanning the QM/MM boundary from 32 to 28. The charge for each QM region, including contributions from both residue protonation state and substrate charge states, ranges from a net charge of +2 for the minimal QM model **1** up to -1 for the largest model **10** (Table 1).

The range of sampled QM regions was chosen in part in order to study the effect of incrementally incorporating residues that complete the $Mg^{2+}$ coordination sphere (axial water,

D141, D169, and N170) or were observed experimentally to have a significant role on catalytic efficiency (E6, W38, Y68, W143, and K144; see Figure 1).[9] Some of these experimentally identified residues may contribute more directly to dynamic effects and structural stability (e.g., the W38 and W143 "gatekeeper"[78] residues that are believed to facilitate substrate binding), whereas our geometry optimizations and reaction pathway analysis should identify residues with the largest electrostatic effect. The $Mg^{2+}$ coordination sphere residues are sequentially incorporated: D141 into model **3** and larger, N170 in model **5** and larger, and D169 in model **8** and larger. Of the experimentally relevant residues, models **2**-**3** include only K144, models **4**-**5** incorporate also Y68, and models **6**-**7** further include W38 and W143 in the QM region. Only the largest models (**8**-**10**) incorporate E6, which forms a hydrogen bond with the SAM-proximal residue Y68. Additionally, up to three water molecules resolved in the active site crystal structure were included in the QM region size sequentially as: one water molecule in models **2**-**4**, two water molecules in model **5**, and all three water molecules in models **6** and larger. None of the external water molecules solvating the protein were included in the QM region, even if they fell within the radial distance cutoff (e.g., for the largest models).

Although most previous QM/MM convergence studies have focused on radial increases in QM region size around an active site,[55,57-59] some alternative schemes have been recently suggested for constructing large QM regions including i) chemical motivation (e.g., hydrogen bonding interactions and close contacts),[54] ii) free energy perturbation analysis,[56] or iii) charge deletion analysis.[50] A motivating factor for radial QM region selection is to avoid biasing QM region choice by incomplete chemical intuition. By sequentially incorporating electronic structure effects from increasingly remote residues, we may identify whether a quantum-mechanical treatment of these residues is required or if a force field description is sufficient. We will also later show that our largest QM region results may be analyzed to determine which residues participate in charge transfer events along the reaction coordinate, permitting identification of the fewest number of QM residues needed for converged QM/MM properties.

A final question is the choice of xc functional to be used. There is now ample evidence that many commonly used xc functionals are poorly suited to large quantum mechanical regions. For example, closure of the highest-occupied/lowest-unoccupied (HOMO-LUMO) gap using semilocal and global hybrid xc functionals has been observed in numerous insulating systems such as polypeptides, proteins, and solvated molecules.[40,102-104] Extending these previous observations, we here find that the QM/MM HOMO-LUMO gap for COMT obtained with global hybrids (e.g., B3LYP[105-107]) closes for QM regions 13 residues and larger (model **4**, see Supporting Information Figure S1). Since a constant, 4 eV gap is maintained for all larger models with ωPBEh,[108] all simulations in this work use this range-separated hybrid (ω=0.5 bohr$^{-1}$).

**4. Results and Discussion**

**4a. QM/MM Convergence of ES Complex Properties**

We now consider the convergence of properties that underlie enzyme catalysis in catechol-*O*-methyltransferase with increasing QM region size in QM/MM simulations. Numerous crystal structures of COMT[89,109-114] have highlighted unusually short SAM methyl to catecholate oxygen (C-O) distances ca. 2.45-2.8 Å in the reactant enzyme-substrate (ES) complex. We carried out structural optimizations, in each case starting from the same 3.11 Å C-O distance well-equilibrated MM structure (see Computational Details) across our 10 different QM region models. We observe significant shortening of the C-O distances over observed values in solution[115] or classical MD,[81-82] particularly when more residues are introduced into the quantum region (Figure 3). In total, C-O distances are reduced by around 0.3 Å with increasing QM region size, from about 3.15 Å in the reactants-only model **1** to 2.85 Å in the largest model **10** with comparable results for model **7** (~500 atoms, 26 residues, a full list is provided in Supporting Information Table S1) and larger. Similar distance reduction is observed for structural optimizations starting from the crystal structure, with distances as short as 2.65 Å favored in the largest QM models (see Supporting Information Figure S2). Differences in results for the two geometry optimizations are likely due to differences in protein structure favored by the MD simulations and the solved X-ray crystal structures, respectively, but the trends are comparable. This reduced distance is consistent with shortened distances in a number of COMT crystal structures,[89,110,113] and is at variance with previous predictions from classical MD treatments[81-82] or quantum mechanical studies with restricted QM regions.[7,83-84]

It is often thought that the accuracy of a given choice for the QM/MM boundary might be affected by the charge state of the QM region (with charge neutrality being preferred) and/or the number of covalent bond cuts connecting the QM and MM regions. In the present work, we do not find a high degree of correlation between these characteristics of the QM region and the accuracy of the resulting QM/MM treatment. For example, changes in the net charge of the QM region cannot explain the variation in distance: the charge differs between models **2** and **3** (from +2 to 0) but the C-O distance continues to decrease for models **5** and **6** where the net charge is 1. In order to assess boundary effects, we computed the minimum distance (min[$d$(link-COM)]) between any link atom and the center of mass (COM) of central SAM (S, C) and catecholate (O) atoms. For all intermediate regions **2-6**, min[$d$(link-COM)] values range from 5.3 Å in model **3** to 6.9 Å in **2**, and this distance lengthens to 7.5-10.0 Å for the largest models (Table 1). In all cases, the proximity of a link atom does not correlate with the distance changes. The largest total number of link atoms closer than 8 Å to the COM ($n_{link}$) is 9 for region **5**, but the optimal C-O distance obtained for this region is in reasonable agreement with the asymptotic limit obtained in larger QM regions. Thus, these results suggest that incorporating more atoms into the QM region does not simply dampen a size or boundary effect. In fact, COMT represents a special case because boundary effects from cutting through covalent bonds in QM/MM regions[17] should be the smallest in the reactants-only model where there are no covalent bonds spanning the QM/MM boundary, and the large reactant size means that most non-minimal model boundaries are distant from the reacting atom COM. Here, our results suggest that specific effects on charge density and polarization of the reactants are only converged when a number of remote residues are treated more flexibly (i.e. quantum mechanically).

As a metric for differences in the substrate electronic structure as QM region size is increased, we evaluated VDD partial charges for these same optimized ES complexes. In isolation, catecholate (CAT) is a singly charged anion; SAM is positively charged with a $S^+$-$CH_3$ moiety as well as a positively charged $NH_3^+$ proximal to a negatively charged terminal carboxylate (see Figure 4 inset). In the smallest model **1**, SAM, CAT and $Mg^{2+}$ (i.e., the entire quantum region) are assigned a total charge of +2, whereas the charge constraint on the active site is relaxed in larger models. The total charge of the SAM and CAT moieties (determined by summing partial VDD charges for atoms in SAM and CAT) slowly approach an asymptotic limit with increasing QM region size (Figure 4) consistent with previous observations for the C-O distance (see Figure 3). SAM partial charges are not monotonic, at first increasing to as much as +1.2 *e* for model **2** from +1.0 *e* for model **1** and then rapidly decreasing to +0.2 *e* in model **3** followed by a slow increase to an asymptotic limit around +0.4 $e$ for model **7** (26 residues, ca. 500 atoms) and larger.

The overall change in charge with QM region size appears to be mediated by charge transfer between the SAM carboxylate and neighboring hydrogen bonding residues (e.g., E90, S72, S119, and H142) that are treated quantum mechanically only in models **3**-**5** and larger (see sec. 4d). This environment stabilizes the donating sulfur, coinciding with an elongation in the $S^+$-$CH_3$ bond by about 0.1 Å. For catecholate, the partial charge changes monotonically with growing QM region size, increasing from -0.25 $e$ in the minimal model **1** to an asymptotic limit of around -0.75 $e$. The increased negative charge on catecholate, as mediated by the surrounding protein environment, would increase electrostatic attraction to the positively charged SAM methyl group and thus promote C-O distance reductions in larger QM models. Overall, our results suggest that the fundamental electronic structure description of the reactants is altered when surrounded by a quantum mechanically-described protein environment rather than a point charge description. The implication of charge transfer also suggests that polarizable force fields would not substantially reduce QM region sensitivity, consistent with some recent observations for polarizable embedding in QM/MM.[31]

**4b. Reaction-coordinate-dependence of QM Region Convergence**

We have shown that the description of the ES complex changes substantially when we increase QM region size. The evolution in electronic structure properties of reactants with increasing quantum mechanical treatment of the protein environment suggests that reactivity may also be modified. In order to confirm this hypothesis, we consider how activation energies and reaction energetics for the rate determining methyl transfer step in COMT vary with increasing QM region size (Figure 5). Methyl transfer activation energies ($E_a$) decrease nearly monotonically from 24 kcal/mol in the minimal model **1** to an asymptotic limit of 16 kcal/mol once 26-30 or more protein residues are included in the QM region. This behavior is similar to the QM region size dependence of C-O distance and CAT/SAM charge shown in Figures 3 and 4. Although we have not incorporated entropic effects here to give a direct comparison to the $\Delta G^{\ddagger}$ (= 18 kcal/mol)[116] obtained from experiments ($k_{cat} \approx 24$/min),[9,116] we would expect the enthalpic barrier to be slightly lower. Thus, we obtain near-quantitative agreement of large-scale QM/MM methyl transfer barriers obtained with range-separated hybrids without need for ad hoc corrections, e.g. due to the use of semi-empirical[7] or semi-local exchange-correlation functionals.[117] As the QM region size is increased, the surrounding QM environment leads to a reduction in reactant distances and charge adjustment on the ES complex that may be viewed as increasing the similarity in the TS and ES structures, leading to a reduced reaction barrier.

The methyl transfer reaction enthalpy ($\Delta E_{Rxn}$) also changes nearly monotonically with increasing QM region size (Figure 5, upper panel), corresponding to increasingly favorable reaction energetics as the QM region is enlarged. Variations in the number of covalent cuts at the boundary or the overall charge of the QM region appear to have little effect and do not correlate with changes in activation energy or reaction energy (Figure 5, lower panel). The minimal model reaction energy is predicted to be weakly endergonic, consistent with previous smaller QM region QM/MM results[86,117-118] on COMT. Instead, the asymptotic limit $\Delta E_{Rxn}$ = -11 to -12 kcal/mol is reached at around model **6** (22 residues). The underestimation of reaction favorability with small QM regions can likely be ascribed to $Mg^{2+}$ coordination: as bidentate catecholate is methylated, its strength as a chelator to $Mg^{2+}$ is weakened. For the small QM regions, $Mg^{2+}$ coordination is mixed between QM and MM residues, with the stabilization by pure MM residue coordination likely insufficient with respect to QM residues because it does not allow for charge transfer. Therefore, the weakening of CAT coordination during methylation is overestimated in smaller QM region models. QM region size impacts both quantitative predictions and qualitative aspects of the COMT methyl transfer mechanism. Currently, free energy barriers computed on the largest QM region sizes studied in this work (ca. 600-1000 atoms) are prohibitive, but we expect that the smaller overall magnitude of entropic contributions to the QM region barrier means that the entropic difference for differing region sizes is likely a substantially smaller contribution than the 8 and 20 kcal/mol differences observed for the activation energy and reaction enthalpy, respectively, from the smallest to the largest QM model.

Although numerous studies have been carried out in evaluating how energetics approach their asymptotic limit with increasingly larger QM-only or QM regions in QM/MM calculations,[50,55-59] none have identified whether geometrical properties of both the reactants and the transition state (TS) converge at similar rates with QM region size. Crystal structures of COMT[89,109-114] all feature unusually short C-O distances, and experimental measurements of kinetic isotope effects[9,82,119] have been suggested by some[120] to be indicative of unusually short C-O or S-O distances in the transition state as well. Here, we identify the transition state approximately as the highest energy structure obtained along the NEB reaction path. In order to compare TS structures for all QM/MM models, we compare both absolute TS geometrical properties, i.e., i) the distance of the methyl donor SAM S to methyl group C, $d$(S-C), and ii) the distance of the methyl acceptor CAT O⁻ to methyl group C, $d$(C-O), as well as relative differences between the TS and the ES complex. Unlike the non-bonded reactant C-O distance (Figure 3), the TS C-O distance shows non-monotonic behavior with increasing QM region size. For QM models **1**-**3**, the C-O distance reduces significantly from ca. 2.0 Å to ca. 1.8 Å (Figure 6). At the same time, the S-C distance lengthens (from 2.3 Å to 2.4 Å), which would lead to identification of a much later transition state if model **3** were used for production calculations. This trend reverses with first lengthening of C-O distances from models **3** to **7** leveling off at around 2.1 Å and a shortening of the S-C distance to about 2.25 Å for the three largest models. In all cases, $d$(S-C) is longer than $d$(C-O) in the TS, but the difference is largest in small QM models and is reduced for the larger models. Key relative geometric properties between the TS and ES complex include SAM S-CAT O$^-$ distance differences:

$$\Delta(\text{S-O}) = d(\text{S-O})|_{\text{TS}} - d(\text{S-O})|_{\text{ES}} \qquad (1)$$

and the lengthening of the S-C bond in the transition state from its equilibrium value:

$$\Delta(\text{S-C}) = d(\text{S-C})|_{\text{TS}} - d(\text{S-C})|_{\text{ES}} \qquad (2)$$

Recall, ES complex $d$(C-O), and thus, $d$(S-O), are monotonically reduced with increasing QM region size (Figure 3). In the TS, the substrate distances, as monitored by $d$(S-O), are even shorter. However, the TS geometry is not affected by QM region enlargement in a manner comparable to the ES complex. Therefore, the relatively large Δ(S-O) of -0.6 Å in the smallest model instead levels off around -0.3 Å for QM/MM models **8**-**10**. That is, large QM treatments impact the TS geometry less, and the enlargement of the QM region causes the ES complex structure to become more transition state-like. This observation is reinforced by Δ(S-C), which also decreases from 0.5 Å to under 0.4 Å. Such a result suggests that the strong dependence of methyl transfer activation energies on QM region size (Figure 5) arises from a lack of cancellation of errors between the ES complex and the TS. These observations reinforce the need to study QM/MM model convergence at multiple points along the reaction coordinate, which has only occasionally been carried out.[50]

Geometric analysis of the ES complex and transition state (TS) has revealed differences in sensitivity to QM region definition. Substrate partial charge analysis in the ES complex (see Figure 4) suggests that QM region sensitivity in COMT is at least somewhat due to charge transfer between the substrates and the surrounding protein. At the transition state, the formal charge on either substrate fragment is likely to be smaller, and thus we investigate whether the TS partial charges show altered QM region size sensitivity compared to the ES complex reactants, R. We quantify the charge transfer from the substrates to the environment in two ways: i) the core substrate partial charge, which is the partial charges summed over both SAM and catecholate and ii) the $Mg^{2+}$ partial charge. Together, i) and ii) must be equal to +2 for our minimal model **1** but could deviate from this idealized value for larger models. Indeed, as the QM region size is increased, both TS and R core and $Mg^{2+}$ partial charges become much more neutral with net overall charges approaching an asymptotic limit around -0.20-0.25 $e$ for the TS and -0.30-0.35 $e$ for R (Figure 7). The partial charge on $Mg^{2+}$ (Figure 7) approaches an asymptotic limit around 0.3 $e$ even more slowly at model **8** but in a similar fashion for both R and TS. This trend in $Mg^{2+}$ charge appears to be derived from inclusion of the $Mg^{2+}$ coordination sphere residues in the QM region.

It is useful to compare whether the difference in the partial charges in the reactant and transition state, Δq(TS-R), converge faster than absolute charges alone through cancellation of errors. For the smallest regions **1**-**2**, the core Δq(TS-R) is constrained by limited region size to be nearly zero, but the TS becomes relatively more positive by up to 0.20 $e$ for intermediate regions **3**-**5**. A loss of 0.1 $e$ from the R core to the TS is observed for models **7** and larger, consistent with the slow approach to a constant value observed in properties of the ES complex alone. The $Mg^{2+}$ Δq(TS-R) similarly approaches a constant value at around -0.025 $e$ for model **8** and larger.

Although the difference in the R and TS $Mg^{2+}$ partial charge is small, it does change sign (e.g., from model **4** to model **5**), indicating high sensitivity to the surrounding environment. From either the perspective of the reacting substrates or $Mg^{2+}$ co-substrate, differences in the electronic environment between the R and TS do not benefit from cancellation of errors, explaining the slow approach to asymptotic limits of reaction energetics.

**4c. Mechanistic Insight from Large-scale QM/MM Reaction Pathway Analysis**

Using the large QM region model **10**, we may identify how substrate partial charges evolve along the methyl transfer reaction coordinate (Figure 8). First, we approximate the reaction coordinate ($\Delta$) by the difference in the SAM sulfur methyl donor distance to the methyl carbon ($d$(S-C)) and the catecholate oxygen methyl acceptor distance to the methyl carbon ($d$(C-O)):

$$\Delta = d(\text{S-C}) - d(\text{C-O}) \qquad (3)$$

Values of $\Delta$ obtained from the model 10 reaction coordinate are provided in Supporting Information Table S2. In order to sum charges along this reaction coordinate, we recall that the methyl group transfers from SAM to catecholate with O-methylated catechol (OMC) and AdoHomocysteine (AdoHcy) as the products. Therefore, we subdivide the partial charge on the methyl group (green open triangles in Figure 8) between the SAM/AdoHcy ($q_S$) and CAT/OMC ($q_C$) fragments according to the relative position, $i$, of the methyl group along the reaction coordinate, $\Delta$, between reactant (R) and product (P) states:

$$q_S^i = q_{\text{AdoHcy}}^i + q_{CH_3}^i \left[ \frac{\Delta(P) - \Delta(i)}{\Delta(P) - \Delta(R)} \right] \text{ and} \qquad (4)$$

$$q_C^i = q_{\text{CAT}}^i + q_{CH_3}^i \left[ \frac{\Delta(i) - \Delta(R)}{\Delta(P) - \Delta(R)} \right] \qquad (5)$$

These partitioned charges and results from alternative partitioning schemes are provided in Supporting Information Table S3. At the highest energy point identified along the reaction coordinate for model **10**, Δ is ca. 0.32 Å, close to the point (Δ=0.40 Å) where the methyl group partial charge is equally divided between $q_S$ and $q_C$. Analysis of the CAT/OMC partial charges at the transition state reveals that the methyl acceptor is still somewhat reactant-like with a negative charge around -0.4 *e* even after including half of the highly partially charged methyl group (+0.4 *e*, the unmethylated CAT partial charges are shown in open circles in Figure 9). Similarly, SAM partial charges remain weakly positive even at the transition state (+0.1 *e*) and not substantially changed with respect to the reactant structure.

Considering even further the close-range interaction between the methyl acceptor on catecholate and the transferring methyl group, electrostatic attraction between the two species (+0.4 *e* for the methyl group, -0.6 *e* for the unmethylated catecholate) is still substantial at the transition state. Later in the reaction coordinate (Δ ca. 0.75-0.9) SAM becomes negatively charged, and the catecholate is neutralized at around -0.1 *e* or less negative only for Δ > 1.0 Å. This range of Δ = 0.75 to 1.0 Å corresponds to *d*(S-C)>2.50 Å and *d*(C-O)<1.75 Å, which is a near product like state both in terms of geometry (in the products, *d*(C-O)=1.44 Å) and energetics. Thus, earlier suggestions[7,86] that electrostatic attraction in the reactants is annihilated at the transition state is apparently an oversimplification when charge transfer is permitted between substrates and the enzyme.

Comparison to a minimal model **1** reveals that constraining the charge on SAM, catecholate, and $Mg^{2+}$ to +2 for the entire reaction will produce a positively charged catecholate acceptor at the transition state (+0.1 *e*, see Supporting Information Figure S3). The net +2 charge in the QM system is distributed over $Mg^{2+}$ (+1.2 *e* in the reactants) and SAM (+1.0 *e* in the reactants) but this leaves little room for CAT to accumulate a strong negative charge (-0.2 *e* in the reactants). Instead, we observed that when this artificial constraint was lifted by enlarging the model, the core substrates (SAM and catecholate) accumulated negative charge from surrounding residues (see Sec 4b). Thus, the electrostatic attraction between reactants is considerably weaker in the minimal model **1** than the large model **10**, which serves as a possible physical origin for the 8 kcal/mol higher barrier in **1** versus **10**. In model **1**, the non-methyl part of CAT becomes positive in the product state (+0.3 *e*, +0.6 *e* with the methyl group), thus making it a very poor chelator to $Mg^{2+}$, also explaining the endothermic reaction energy observed earlier for the minimal model (compared to exothermic reaction energy for the large model, as discussed in Sec. 4b). These results suggest that the enzyme, and $Mg^{2+}$ in particular, mediates charge transfer between the reactants and the environment, extending the portion of the reaction coordinate over which electrostatic attraction between the two fragments is favorable past the transition state.

## 4d. Obtaining Atom-Economical QM Regions

Following confirmation that key properties of the COMT enzyme are consistent for large radially-cut QM regions (ca. 600-1000 atoms) in QM/MM calculations, we now aim to identify the subset of residues included in these QM regions that impact reaction coordinate properties most strongly. COMT is a challenging system for QM/MM convergence studies because the SAM, catecholate, and $Mg^{2+}$ substrates alone span a large portion of the protein's solvent-exposed active site, and residues proximal to one substrate may be distant from another. As noted previously (see Figure 7), the core substrates (SAM, catecholate, and $Mg^{2+}$) carry more negative partial charge than expected from nominal charge assignment, and the total charge evolves as the reaction progresses. Therefore, we first identify which residues have a variation in total electron density during the methyl transfer reaction and are thus acting as charge sources or sinks for the substrates. Using the largest model **10** holoenzyme studied in this work, we computed the per-residue VDD partial charge sums, i.e.:

$$q_{res}^{VDD} = \sum_{j \in res} q_j^{VDD} \qquad (6)$$

on each residue in the QM region for 21 snapshots interpolated along the methyl transfer reaction coordinate (see Supporting Information Tables S4-S7). Total charges of some residues appear to vary strongly with the reaction coordinate, e.g. M40 and N41, which accumulate around -0.15-0.2 *e* over the course of the reaction coordinate (Supporting Information Figure S4). However, the partial charge of most residues fluctuates across the methyl transfer coordinate, and there is limited correlation between variation and relative proximity to the substrates in the active site (see Supporting Information Figure S5).

In order to isolate the charge fluctuations most relevant to the substrate environment, we removed SAM, catecholate, and $Mg^{2+}$ from each snapshot and repeated the summed-over-residue VDD computations (see Supporting Information Tables S8-S11). In both cases, the sum was computed two ways: with link atoms assigned to their respective residue or excluded, and the no link atom data was used here due to lower fluctuations observed in the following analysis (i.e., link-atom-derived charge fluctuations may lead to false positives, see Supporting Information). The residues that display the largest holo-apo charge shift are expected to be essential to the complete description of the electronic environment in the active site, as a point-charge electrostatic description afforded by MM alone should be insufficient. In order to quantify and rank importance of residues by their interactions with substrates, we compute the difference between the apo and holo residue-summed partial charges ($q_{res}$) and average them over the reaction coordinate as follows:

$$\Delta q_{\text{res,av}} = \frac{\sum_{i}^{n} q_{res,i}^{\text{VDD,apo}} - q_{res,i}^{\text{VDD,holo}}}{n} \tag{7}$$

Thus, residues that lose charge when the substrates are removed have a negative $\Delta q_{\text{res,av}}$, whereas ones that gain charge back have a positive $\Delta q_{\text{res,av}}$.

In total, we find 11 residues with $\Delta q_{\text{res,av}}$ at least 0.05 $e$ in magnitude (Figure 9). These 11 residues include i) hydrogen bond acceptors to SAM (E90, E64, H142) and catecholate (E199) that lose substantial charge, ii) hydrogen bond donors to SAM (S72), iii) $Mg^{2+}$ coordination sphere residues that alternately lose (N170, D169) or gain charge (D141), and iv) a cluster of residues behind the SAM substrate that forms more indirect interactions (V42, A67, A73). Residues in cases i-iii would have likely been identified with the help of chemical intuition, but other residues that may have been deemed important through chemical intuition arguments alone, e.g. the catechol deprotonating K144, are absent from this list. Similarly, residues that may be key to substrate binding and protein dynamics (e.g., gatekeeper residues W38 and W143) do not impact charge on the substrate and therefore are also not detected by this analysis. Alternatively, proximity may have been a useful strategy for identifying V42, A67, and A73 as relevant residues, since the three residues are adjacent to SAM, but several residues proximal to CAT (e.g., K144, P174, L198) do not show comparable charge sensitivity. The shape of the space occupied by residues with large charge shifts (white and blue sticks shown in inset in Figure 10) is still centered on the substrates but ellipsoidal in nature. Thus, although several of the residues are included in our 2$^{nd}$ or 3$^{rd}$ smallest radial QM regions (V42, E90, D141, N170, E199), others coincide with the intermediate models **4**-**6** (S72, A67, H142), and still others only appear in the larger model **8** (E64, D169, A73) (see Supporting Information Table S12).

An additional 5 residues (M40, N41, Y68, I91, S119, shown as green sticks in Figure 10 inset) have single-snapshot $\Delta q_{res}$ that meet or exceed 0.05 $e$ in magnitude for at least one snapshot (see Supporting Information Tables S9 and S12). The $Mg^{2+}$- and catecholate-adjacent residues M40 and N41 have negligible $\Delta q_{res}$ in the first half of the reaction, but values increase at the transition state and towards the products. Conversely, the Y68 residue, which has been the focus of previous experimental and computational mutagenesis efforts,[9,82] has a large charge shift in the first portion of the reaction but limited effect after the transition state. These five additional residues are present in our original models **3**-**5**. Now, we identify if the results of our charge shift analysis can be used to prune or refine large radial cuts of QM regions in analogy to charge deletion analysis.[50,121-122] Charge deletion analysis has been used[50] with the assumption that any strong QM-MM electrostatic interaction cannot be properly accounted for across the QM/MM boundary, favoring placing that residue in the QM region.[50] Any proximal MM residue with moderately strong point charges will be identified by charge deletion analysis, but we wish to take the more economical view that some QM-point charge interactions are in fact suitably treated with QM/MM. Thus, we hypothesize that adequate QM regions may instead be constructed on the basis of the residues that exhibit large charge shifts in response to the substrates.

We now construct new QM models from the residues identified in charge shift analysis and compare to the original radial models. Both the 11 residue (214 atoms) and enlarged 16 residue (296 atoms) models are similar in size to model **4** (13 residues, 268 atoms) but are comprised of different residues. Both of these new models are substantially smaller than the models (**7-8**, 26-34 residues, 497-600 atoms) we previously identified as consistent with the largest model **10** across all properties considered in this work. The 16 residue model omits K144, which appears in our radial models **2** and larger, as well as G66 and Y71, which both appear in radial model **4** (Supporting Information Table S13). All three models have comparable distance (6.4-6.7 Å) between the closest link atom and the center of mass of the S-C-O bond. The new 11 and 16 residue models, which we will refer to as **4A** and **4B**, respectively, reduce the total number of link atoms closer than 8 Å to the S-C-O bond center of mass from 7 in model **4** to 5. The total number of link atoms for **4A** (16) and **4B** (18) are reduced slightly as well from model **4** (20). The omission of K144 and inclusion of a number of anionic residues, however, imparts the largest net negative charge to both of these regions (-3) compared to any of the previous radial models (-1).

We computed partial charges and reaction pathways for these new models and compare both to model **4** as a reference for equivalent computational cost as well as to the largest QM/MM model **10** (Table 2). The root sum squared (RSS) error of evaluated residue-summed partial charges ($q$) for model M with respect to the reference model **10** is evaluated as:

$$\text{RSS}(q,\text{M}) = \sqrt{\sum_{\text{res}} (q_{\text{res,M}} - q_{\text{res},10})^2} \tag{8}$$

In total, we evaluate the i) reactant (SAM, catecholate, and $Mg^{2+}$), ii) transition state (the reactants with $CH_3$ partitioned as described in sec. 4c), and iii) product (AdoHcy, OMC, and $Mg^{2+}$) partial charges summed over each residue for a total of 9 terms in the sum in eqn. 8. As suggested by secs. 4a-4b, model **4** RSS partial charge error is quite large at 0.7 due to increased partial positive charge on $Mg^{2+}$, reduced charge separation in the transition state, and enhanced charge separation in the products. The 11-residue and 16-residue models **4A** and **4B**, on the other hand, have good and near quantitative agreement in partial charges with the larger model **10** with an RSS of 0.3 and 0.0, respectively. Disagreement for the smaller model is primarily due to increased negative charge on SAM across the reaction coordinate.

We also converged methyl transfer pathways for the **4A/4B** models and compute the RSS error in the activation and reaction energies as:

$$\text{RSS}(E, \text{M}) = \sqrt{\left(E_a^{\text{M}} - E_a^{10}\right)^2 + \left(\Delta E_{\text{rxn}}^{\text{M}} - \Delta E_{\text{rxn}}^{10}\right)^2} \quad (9)$$

Using this metric, the radial model **4** has a 7.2 kcal/mol RSS error due to overestimating the barrier height and underestimating reaction exothermicity. In contrast, model **4B** yields near-quantitative agreement of 0.4 kcal/mol RSS error due to sub-kcal/mol differences in barrier height and reaction energetics, and the smaller model **4A** is also in very good agreement with an RSS of 1.6 kcal/mol (Table 2). Comparison of the full methyl transfer reaction profiles (Figure 10) reveals that model **4B** overlaps nearly exactly with model **10**, whereas model **4A** shows a slightly earlier transition state with lower barrier and less exothermic products. However, neither show the large deviations apparent between the comparably sized model **4** and the large-scale model **10**, where the full reaction profile highlights again differences in the character of the much later transition state structure as well as qualitative differences in barrier height and shape. Thus, properties consistent with large-QM/MM models may be obtained at a fraction of the computational cost from QM regions with as few as 214-296 atoms as long as the optimal QM residues are selected.

Based on these promising results, we propose a general protocol for unbiased QM region determination in QM/MM calculations: i) partial charges (or other relevant properties[54]) of reacting substrates should be obtained from very large radial models that have no link atoms adjacent to the central active site in a reactant, product, and key intermediate or transition-state-like geometries, ii) the calculations should be repeated with the reacting substrates removed. If the substrates or catalytic center are covalently linked to the protein, mutagenesis rather than complete substrate removal may be necessary, and iii) the residues for which there is an apparent significant charge or property difference (e.g., 0.05 $e$ or greater difference in charge) from i or ii should be used to construct a new QM region for QM/MM calculations. iv) This new QM/MM model may be validated through energetic, structural, or partial charge properties for agreement with the large radial model results. In total, this charge shift analysis requires no more than a handful of very large (ca. 1000 QM atoms) QM/MM calculations and facilitates a systematic and unbiased determination of an atom-economical QM region that will not require strong chemical intuition nor potentially overestimate electrostatic interactions that are suitably treated across the QM/MM boundary.

## 5. Conclusions

We have quantified how key descriptive properties of enzyme catalysis obtained from simulations depend on the size of the QM region in QM/MM calculations for an enzyme in which the smallest possible QM regions do not suffer from boundary effects. Our results on COMT show that geometric and electronic structure properties of the reactants are slow to approach asymptotic limits as remote residues are added radially to the QM region. Namely, both reactant distances and partial charges on reactants converge slowly with increasing QM region size. By carrying out extensive geometry optimizations and transition state searches carried out with range-separated hybrid DFT made possible through GPU-accelerated quantum chemistry, we have separated substrate property convergence with increasing QM region size from well-known errors of semi-local exchange treatments in large QM system sizes.

For the COMT example investigated here, radial QM region models that are ≈10x larger than typically used in QM/MM calculations are needed for consistent structural properties or reaction energetics. Although properties such as forces on central QM atoms have been shown to converge only when QM regions were at least 500 atoms in size,[46] it remained possible that error cancellation along a reaction coordinate might instead lead to good prediction of relative properties with smaller QM region size. We have instead demonstrated that differences in TS and ES complex property convergence lead to poor cancellation of errors due to differences in charge transfer and residue interactions along the reaction coordinate.

Using our large QM/MM models, we also provided mechanistic insight into the role of the enzyme environment on methyl transfer. Namely, we observed that charge annihilation between the oppositely charged reactants does not occur until after the transition state structure and that the charge transfer between substrates and the protein environment primes the ES complex to be more TS-like.

Finally, we introduced charge shift analysis to pare down large QM models into a minimal set of residues needed for quantitative accuracy. By incorporating only the residues that participated in charge transfer with the reactants, quantitative agreement with a 56 residue (968 atom) radial QM region was reached with only 16 residues (296 atoms). Although our analysis revealed several residues that might be selected on the basis of chemical intuition or proximity to substrates, other residues that would have been selected under either criterion were identified to be unnecessary. Several nonpolar residues that would have escaped selection using typical criteria were identified as important. Future work will be aimed toward validating this and related approaches for unbiased, automated determination of optimal QM regions in QM/MM calculations across a range of enzyme classes.

## ASSOCIATED CONTENT

**Supporting Information Available:** Details of all QM region model sizes employed; C-O distance region dependence for QM/MM optimizations of 3BWM crystal structure; individual distances along reaction coordinate; partial charge sums over core substrate definitions along reaction coordinate for largest and smallest models; sums per residue with and without link atoms for large QM model; apo-holo charge sum differences with and without link atoms; charge variation plot along reaction coordinate; heat map of residue absolute charge fluctuations; zip file containing PDB coordinates of initial reactant structure, reactants, transition states, and products as well as pymol session file for visualizing QM regions. This material is available free of charge via the Internet at http://pubs.acs.org.

## AUTHOR INFORMATION

**Corresponding Author**

*email: todd.martinez@stanford.edu

**Notes**

The authors declare the following competing financial interest(s): T.J.M. is a cofounder of PetaChem, LLC.

## ACKNOWLEDGMENT

This work was supported by the Department of Defense (Office of the Director of Defense Research and Engineering) through a National Security Science and Engineering Faculty Fellowship (to T.J.M) and by National Institutes of Health (NIH) grants to J.P.K. and J.Z.

(GM025765 and GM039296). H.J.K. holds a Career Award at the Scientific Interface from the Burroughs Wellcome Fund.Kulik, et al. – QM Region in COMT – Page 28

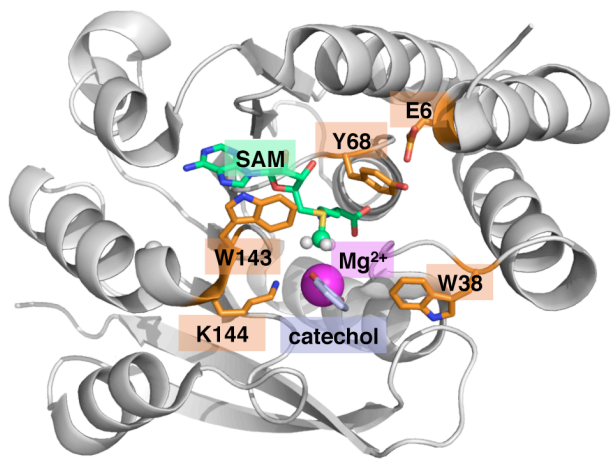

**Figure 1.** COMT protein with active site features highlighted. The reactants (SAM and catechol) are shown in green and purple, respectively, as well as an $Mg^{2+}$ ion in magenta and five key residues identified from experiments (orange).

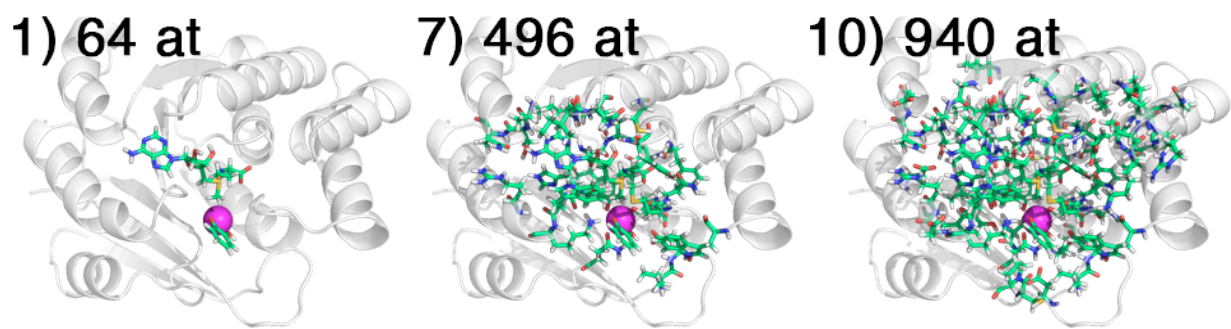

**Figure 2.** QM regions shown for models 1 (64 atoms, 0 residues), 7 (496 atoms, 26 residues), and 10 (940 atoms, 56 residues) with QM atoms shown in green stick representation.

**Table 1.** Summary of QM regions studied in this work. Quantities include the model number, radius of the cut used to define the region, number of non-substrate or cofactor residues, number of QM atoms, number of link atoms, total number of atoms in the QM calculation (link atoms plus QM atoms), charge assigned to the QM region, the minimum distance between the central reacting atoms (S, C, O) center of mass (COM) and the closest link atom(min[$d$(link-COM)]), and the number of link atoms within 8 Å ($n_{link}$<8Å) of the COM.

| Region | radius (Å) | # res. | # QM atoms | # link atoms | total atoms | QM charge | min[$d$(link-COM)] (Å) | $n_{link}$<8Å |
|---|---|---|---|---|---|---|---|---|
| 1 | 0.00 | 0 | 64 | 0 | 64 | +2 | -- | -- |
| 2 | 1.75 | 3 | 120 | 6 | 126 | +2 | 6.9 | 1 |
| 3 | 2.00 | 7 | 172 | 14 | 186 | 0 | 5.3 | 3 |
| 4 | 2.25 | 13 | 268 | 20 | 288 | 0 | 6.7 | 7 |
| 5 | 2.50 | 19 | 387 | 26 | 413 | +1 | 6.0 | 9 |
| 6 | 2.75 | 22 | 448 | 24 | 472 | +1 | 5.5 | 7 |
| 7 | 3.00 | 26 | 497 | 24 | 521 | +1 | 7.5 | 5 |
| 8 | 5.00 | 34 | 600 | 32 | 632 | -1 | 7.5 | 4 |
| 9 | 6.00 | 43 | 738 | 28 | 766 | -1 | 7.5 | 2 |
| 10 | 7.00 | 56 | 940 | 28 | 968 | -1 | 10.0 | 0 |

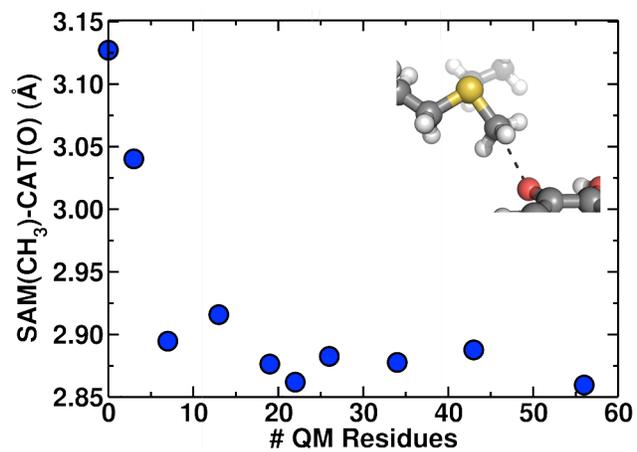

**Figure 3.** Dependence of reactant distances in the protein (distance from transferring methyl carbon of SAM to acceptor oxygen of catecholate) on QM region size. QM region sizes are reported in terms of the number of protein residues included in each QM region from reactants-only (0 residues, 64 atoms) to a 7 Å radius around the reactants (56 residues, 940 atoms). Inset shows the orientation of the methyl donor and acceptor.

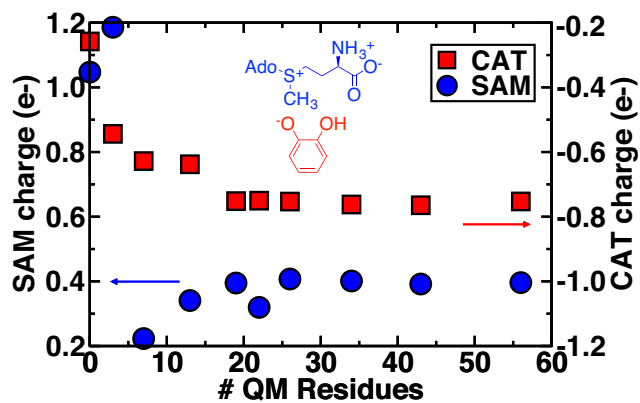

**Figure 4.** Dependence of partial charges for reactants on QM region size. Charges on SAM (blue circles, blue structure) and catecholate (CAT, red squares, red structure) are compared with the values indicated on left and right y-axes, respectively. The scale of the y-axis is the same for SAM and CAT but the charges are of opposite sign (positive for SAM and negative for CAT).

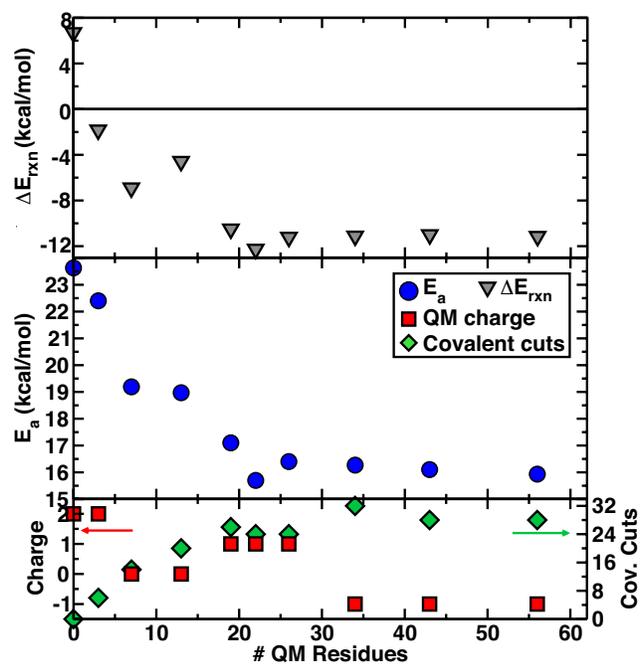

**Figure 5.** Dependence of methyl transfer activation energy on QM region size (top) compared to variation in charge of QM region (red squares) and number of covalent cuts in QM region (green diamonds) with increasing region size.

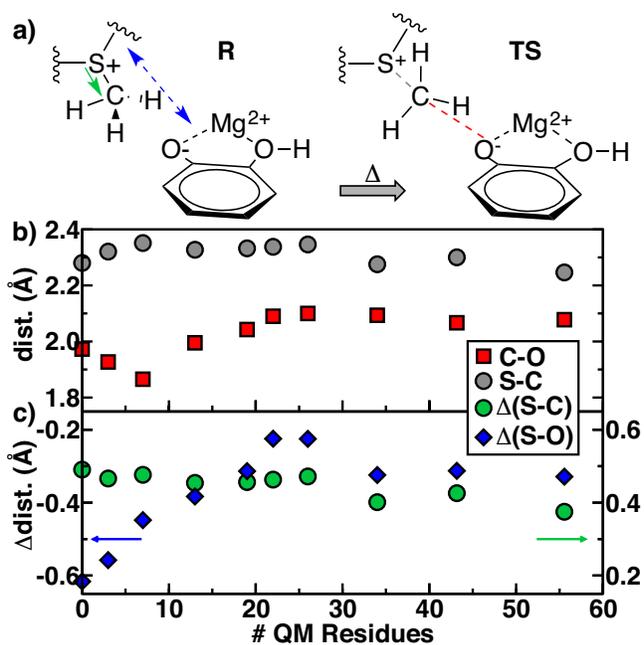

**Figure 6.** a) Reactant (R) and transition state (TS) structures annotated with TS-R ΔS-C distance and TS-R ΔS-O distance (left) and S-C or C-O distance (right). b) S-C (gray circles) and C-O (red squares) distances with QM region size. c) shift from R to TS of S-C (green circles) and S-O (blue diamonds) distances with QM region size.

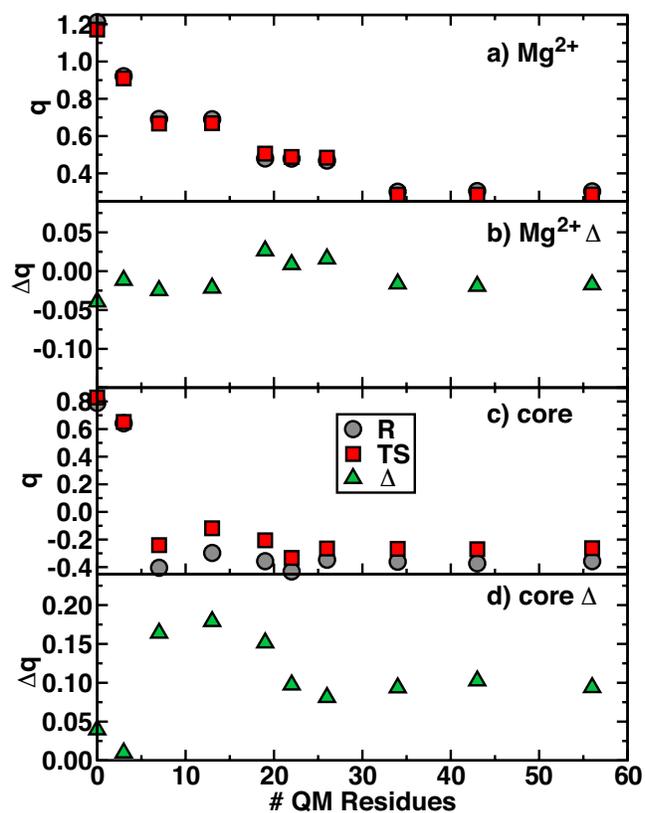

**Figure 7.** a) Reactant (R) and transition state (TS) partial charges and b) TS-R partial charge differences for $Mg^{2+}$. c) R and TS partial charges summed over the core (SAM and catecholate only) and d) TS-R differences for the core.

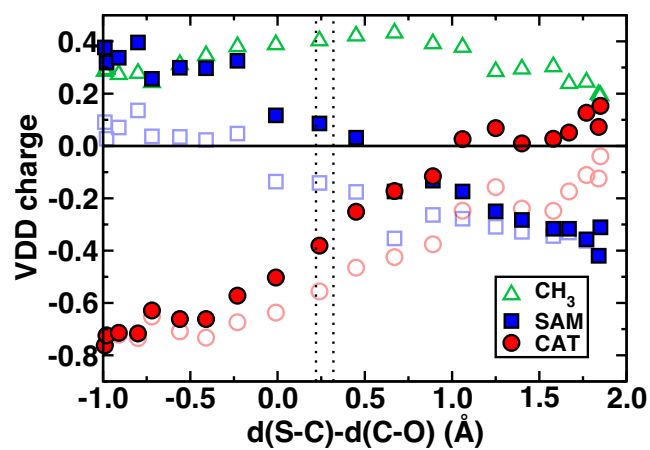

**Figure 8.** By-residue partial charges of SAM (blue filled squares) and catecholate (CAT, red filled circles) along the reaction coordinate defined by the difference in the distance of the transferring methyl carbon to the donor SAM S atom and the acceptor catecholate O atom, as described in the main text. The transition state region is shown as two vertical dotted lines. For comparison, sums of the charge over the methyl group only (green open triangles), adohomocysteine (blue open squares), and unmethylated catecholate (red open circles) are also shown.

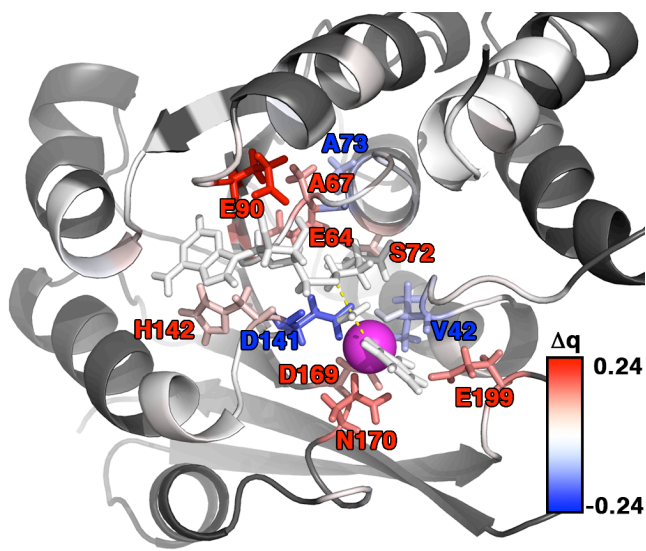

**Figure 9.** Difference of by-residue VDD charge sums upon removal of the substrates (SAM, catecholate, and $Mg^{2+}$) with substrates in transition state structure shown in white sticks. Residues shown and labeled in blue lose partial charge upon substrate removal, whereas residues shown and labeled in red gain partial charge upon substrate removal. All residues with $\Delta q \geq |0.05|$ are shown as sticks. All remaining QM residues are shown in cartoon as white, whereas MM residues are shown in dark gray.

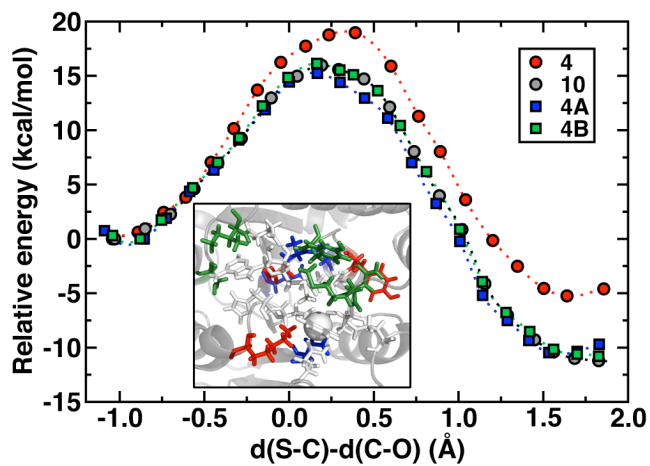

**Figure 10.** Methyl transfer reaction profiles for original models **4** and **10** (red and gray circles, respectively) and optimized models **4A** and **4B** (blue and green squares, respectively). Inset shows atoms in all QM region models (white sticks), only in model **4** (red sticks), added in model **4A** or **4B** (blue sticks), or added in model **4B** (green sticks). For model **10** QM region, see Figure 2.

**Table 2.** Properties, (number of residues, #res, number of QM atoms #at, number of link atoms, # link, closest link atom (in Å), and number of link atoms, n, within 8 Å), partial charges, (for reactant, R, transition state, TS, and product, P, SAM, CAT and Mg), energetics, and errors (activation energy, $E_a$ and reaction energy, $\Delta E_{rxn}$), and root sum squared (RSS) errors in charge and energy of new models 4A and 4B alongside model 4 as determined by agreement with the largest model 10.

| M | Region Properties | | | | | R charges | | | TS charges | | | P charges | | | Energetics (kcal/mol) | | RSS errors | |
|---|---|---|---|---|---|---|---|---|---|---|---|---|---|---|---|---|---|---|
| | # res | # at | # link | closest | n < 8 | SAM | CAT | Mg | SAM | CAT | Mg | SAM | CAT | Mg | $E_a$ | $\Delta E_{rxn}$ | Charge | Energy |
| 4A | 11 | 214 | 16 | 6.4 | 5 | 0.22 | -0.67 | 0.31 | -0.01 | -0.33 | 0.32 | -0.62 | 0.09 | 0.35 | 15.2 | -9.7 | 0.3 | 1.6 |
| 4 | 13 | 268 | 20 | 6.7 | 7 | 0.34 | -0.64 | 0.69 | 0.09 | -0.21 | 0.67 | -0.76 | 0.22 | 0.53 | 18.9 | -4.6 | 0.7 | 7.2 |
| 4B | 16 | 296 | 18 | 6.6 | 5 | 0.38 | -0.77 | 0.31 | 0.13 | -0.41 | 0.29 | -0.44 | 0.10 | 0.34 | 16.2 | -10.8 | 0.0 | 0.4 |
| 10 | 56 | 940 | 28 | 10 | 0 | 0.38 | -0.76 | 0.30 | 0.12 | -0.38 | 0.29 | -0.42 | 0.07 | 0.34 | 15.9 | -11.2 | 0.0 | 0.0 |

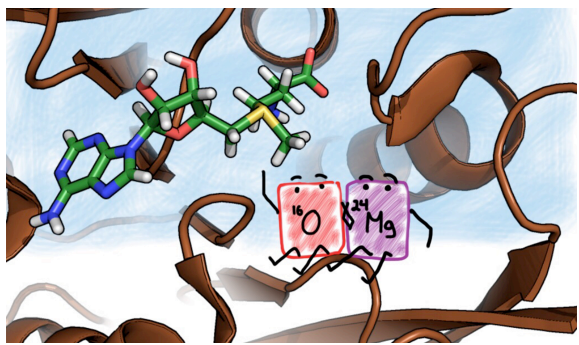